\documentclass[aps,reprint,prl,superscriptaddress]{revtex4-1}

\pdfoutput=1

\usepackage{bm}
\usepackage{amssymb}
\usepackage{amsmath}
\usepackage{graphicx,color}
\usepackage{soul}
\usepackage[normalem]{ulem}

\begin{document}

\title{Grain Boundary Structures and Collective Dynamics of Inversion Domains
  in Binary Two-Dimensional Materials}
\author{Doaa Taha}
\affiliation{Department of Physics and Astronomy, Wayne State University,
Detroit, Michigan 48201, USA}
\author{S. K. Mkhonta}
\affiliation{Department of Physics and Astronomy, Wayne State University,
Detroit, Michigan 48201, USA}
\affiliation{Department of Physics, University of Swaziland, Private Bag 4,
Kwaluseni M201, Swaziland}
\author{K. R. Elder}
\affiliation{Department of Physics, Oakland University, Rochester, Michigan
48309, USA}
\author{Zhi-Feng Huang}
\affiliation{Department of Physics and Astronomy, Wayne State University,
Detroit, Michigan 48201, USA}

\begin{abstract}
  Understanding and controlling the properties and dynamics of topological defects
  is a lasting challenge in the study of two-dimensional materials, and is crucial 
  to achieve high-quality films required for technological
  applications. Here grain boundary structures, energies, and dynamics
  of binary two-dimensional materials are investigated through the development
  of a phase field crystal model that is parameterized to match the ordering, 
  symmetry, energy and length scales of hexagonal boron nitride.
  Our studies reveal some new dislocation core structures for various symmetrically
  and asymmetrically tilted grain boundaries, in addition to those obtained in
  previous experiments and first-principles calculations. We also identify a
  defect-mediated growth dynamics for inversion domains governed by the collective
  atomic migration and defect core transformation at grain boundaries and junctions,
  a process that is related to inversion symmetry breaking in binary lattice.
\end{abstract}

\maketitle

Topological defects, such as dislocations and grain boundaries (GBs), are
known to be pivotal in controlling material properties. It is challenging
to effectively capture the complexity of defects, given the nonequilibrium nature
of material growth and evolution processes. Recent progress in the study of
two-dimensional (2D) hexagonal materials, such as graphene, hexagonal boron
nitride (h-BN), and transition metal dichalcogenides (TMDs), provides an excellent
platform for the investigation of defect properties and dynamics. This is driven
by the demand for controllable fabrication and synthesis of large-scale,
high-quality samples of these atomically thin systems, which mostly rely on
vapor-phase heteroepitaxy techniques particularly chemical vapor deposition.
Such large-area 2D epitaxial films are usually polycrystalline \cite{Yazyev14},
with various types of defects found in both theoretical
\cite{Yazyev10,LiuNanoLett10,LiuACSNano12,ZouNanoLett13} and experimental
\cite{Huang11,GibbJACS13,CretuNanoLett14,LiNanoLett15,vanderZande13,BjorkmanSciRep13}
studies of 2D materials. Typical examples include penta-hepta ($5|7$)
defects in graphene \cite{Huang11} and either penta-hepta or square-octagon
($4|8$) defects in h-BN \cite{GibbJACS13,CretuNanoLett14,LiNanoLett15} and TMD
\cite{vanderZande13} sheets.

Compared to 2D single-component materials such as graphene,
in binary hexagonal materials (e.g., h-BN and TMDs)
the inversion symmetry is broken in the corresponding binary honeycomb lattice.
A much richer variety of GB configurations can be identified, some of which
can significantly alter system electronic properties, as predicted by
first-principles calculations \cite{LiuACSNano12,ZouNanoLett13} and
found in experiments of h-BN \cite{LiNanoLett15}, MoS$_2$
\cite{vanderZande13,ZhouNanoLett13} and MoSe$_2$ \cite{LiuPRL14}
epitaxial monolayers. Of particular interest are the $60^{\circ}$ GBs
[i.e., inversion domain boundaries (IDBs)], a characteristic of inversion
symmetry breaking. Depending on the detailed structures of dislocation
cores, these $60^{\circ}$ boundaries can cause a reduction of band gap
(as in h-BN \cite{LiuACSNano12}), the appearance of mid-gap states (for GBs
consisting of $4|8$ cores in MoS$_2$ \cite{ZouNanoLett13,vanderZande13}),
or a transition from semiconducting to localized metallic modes (for
$4|4$ or $8|8$ cores in MoS$_2$ \cite{ZouNanoLett13,ZhouNanoLett13} and
MoSe$_2$ \cite{LiuPRL14}).

It is of great difficulty to effectively
track or control the dynamics of defect formation over the relevant spatial
and temporal scales, via either \textit{in situ} experimental techniques or
simulations. Experimentally the studies of defect dynamics mostly
rely on the activation process of electron irradiation that generates migrating
vacancies in the sample \cite{CretuNanoLett14,LinACSNano15,LehtinenACSNano15}.
Most theoretical studies are based on atomistic methods
particularly first-principles density functional theory (DFT) and molecular
dynamics. While large progress has been made for identifying
lowest-energy defect structures and their electronic properties
\cite{Yazyev14,Yazyev10,LiuNanoLett10,LiuACSNano12,ZouNanoLett13},
the atomistic techniques are usually limited by the restrictions of small length
and time scales and the pre-constructed defect core configurations. It is thus
important to develop and apply modeling methods that are able to access
large system sizes and realistic time scales while still maintaining
microscopic spatial resolution.

Here we first construct such a model for binary 2D materials, based
on the phase field crystal (PFC) method \cite{Elder02,*Elder04,Elder07,
Emmerich12,Mkhonta13} which can simultaneously address mesoscopic structural
profiles at diffusive time scales and resolve crystalline microstructures
\cite{Huang08,Mellenthin08,TothPRL11,SchwalbachPRE13,AdlandPRL13,
OpokuPRB13,BackofenActaMater14,HirvonenPRB16,Mkhonta16,Huang16}.
For simplicity, here a 2D planar model is developed that does not include
out-of-plane deformations, given the constraint of 2D monolayers during
epitaxial growth. As described in Supplemental Material \cite{SM}, the
model parameters are chosen to match the symmetry, sublattice
ordering, Young's modulus, and atomic spacing of h-BN.
This model is used to systematically study
GB structures, energies and the spontaneous formation of defects in systems
up to micron size, without any pre-determined setup of defect cores.
This allows us to identify complex defect structures that are absent in
previous research, and also to predict a growth mechanism of collective
dynamics and boundary defect shape transformation for inversion domains.

Our model is motivated from the classical dynamic DFT, similar to the previous
procedure for PFC models \cite{Elder07,Huang10}. For a binary AB system, the
dimensionless fields $n_A$ and $n_B$ of atomic number density variation are
governed by the relaxational dynamics, i.e., 
\begin{equation}
\partial n_A / \partial t = \nabla^2 \frac{\delta \mathcal{F}}{\delta n_A}, \qquad
\partial n_B / \partial t = m_B \nabla^2 \frac{\delta \mathcal{F}}{\delta n_B},
\label{eq:nA_nB}
\end{equation}
where $m_B=M_B/M_A$ with $M_{A(B)}$ the mobility of A(B) component, and the 
free energy functional $\mathcal{F}$ is given by
\begin{eqnarray}
  &\mathcal{F}& = \int d{\bm r}  \left [ -\frac{1}{2} \epsilon_A n_A^2
    + \frac{1}{2} n_A \left ( \nabla^2 + q_A^2 \right )^2 n_A 
    - \frac{1}{3} g_A n_A^3 \right. \nonumber\\
  && + \frac{1}{4} n_A^4 -\frac{1}{2} \epsilon_B n_B^2
     + \frac{\beta_B}{2} n_B \left ( \nabla^2 + q_B^2 \right )^2 n_B 
     - \frac{1}{3} g_B n_B^3 \nonumber\\
  && + \frac{1}{4} v n_B^4 + \alpha_{AB} n_A n_B
     + \beta_{AB} n_A \left ( \nabla^2 + q_{AB}^2 \right )^2 n_B \nonumber\\
  && \left. + \frac{1}{2} w n_A^2 n_B + \frac{1}{2} u n_A n_B^2 \right ].
\label{eq:F}
\end{eqnarray}
Here the rescaled parameters are chosen as
$\epsilon_A=\epsilon_B=0.3$, $q_A=q_B=q_{AB}=1$, $\alpha_{AB}=0.5$,
$\beta_{AB}=0.02$, $g_A=g_B=0.5$, $w=u=0.3$, and $\beta_B=v=m_B=1$.
The corresponding stability and phase diagrams are given in Fig. S2 of
Supplemental Material \cite{SM}. We parameterize the model within the binary
honeycomb regime of the phase diagram and match to the energy and length scales
of h-BN. Given the Young's modulus $Y=810$ GPa \cite{KudinPRB01}, lattice
constant $a_0=2.51$ {\AA}, and vertical layer spacing 3.33 {\AA} for h-BN, a
length unit ($l=1$) of our model corresponds to 0.342 {\AA} and a PFC energy
unit corresponds to 2.74 eV. Two pivotal factors have been incorporated in our
model construction, including the lattice symmetry and a factor
specific for binary compounds, that is, the heteroelemental A-B neighboring
should be energetically favored as compared to homoelemental A-A or B-B ones.

\begin{figure}
\centerline{\includegraphics[width=0.43\textwidth]{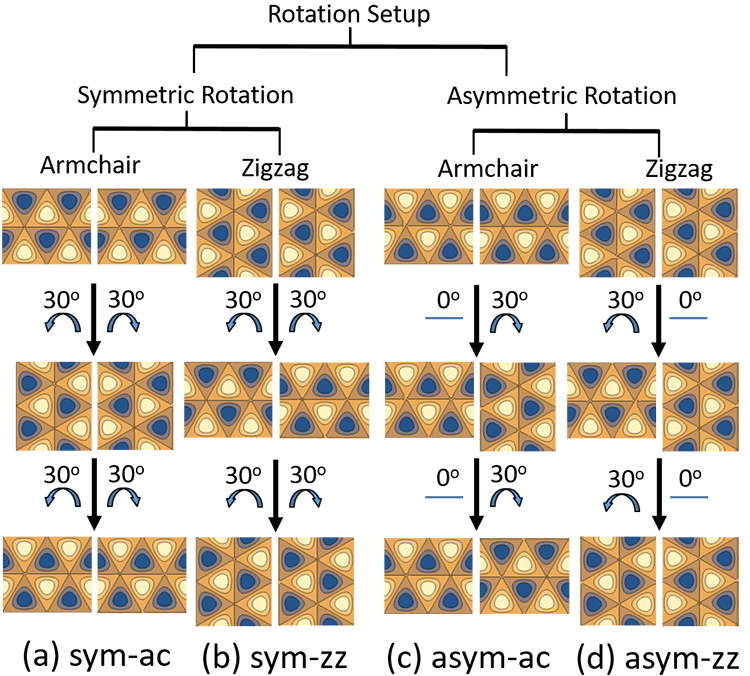}}
\caption{Schematic of initial setups for GBs.}
\label{fig:gbrotation}
\end{figure}

In our PFC study four types of GBs are examined (see Fig. \ref{fig:gbrotation}),
for two adjoining grains of orientations $\theta_1$ and $\theta_2$ with GB
misorientation angle $\theta=\theta_1+\theta_2$. The first two types correspond
to symmetrically tilted GBs with $\theta_1=\theta_2=\theta/2$,
starting either from the armchair edge (Fig. \ref{fig:gbrotation}a, sym-ac,
keeping mirror symmetry between the two grains), or from the zigzag direction
(Fig. \ref{fig:gbrotation}b, sym-zz). Given that polycrystalline samples are
characterized by a prevalence of asymmetrically
tilted GBs (i.e., $\theta_1 \neq \theta_2$), we also simulate the configurations
in the limit of completely asymmetric GBs with $\theta_1=0$ and $\theta_2=\theta$,
for grain rotation from both armchair (Fig. \ref{fig:gbrotation}c,
asym-ac) and zigzag (Fig. \ref{fig:gbrotation}d, asym-zz) directions.

\begin{figure}
\centerline{\includegraphics[width=0.5\textwidth]{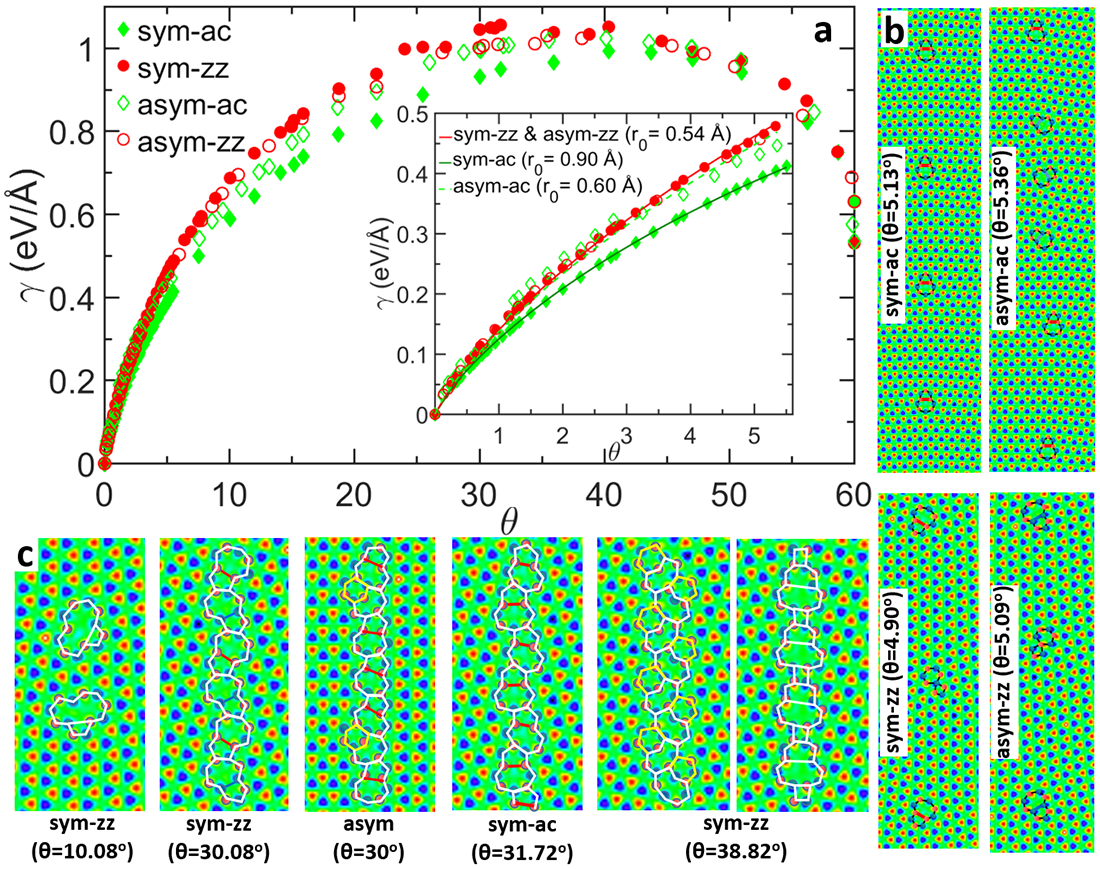}}
\caption{(a) GB energy $\gamma$ as a function of $\theta$.
  Inset: The fitting to the Read-Shockley equation at small angles.
  Also given are sample GB structures for (b) small $\theta$
  and (c) larger angles, with A atomic sites shown in red while B in blue.}
\label{fig:gbenergies}
\end{figure}

The system simulated contains two parallel GBs to satisfy periodic boundary
conditions. The total system size ranges from 14.1 nm $\times$ 24.4 nm for large
misorientations to 0.9 $\mu$m $\times$ 0.3 $\mu$m for very small $\theta$
(corresponding to $1.3 \times 10^4$ to $9.1 \times 10^6$ equivalent atomic sites).
For each angle the two misoriented grains are connected
initially via a narrow band (around 2-6 grid points) of supersaturated 
homogeneous phase which spontaneously solidifies, causing the grains to merge
and form a GB that evolves to a steady state. We tested various connection
conditions by varying the relative lattice translation of adjoining grains to 
obtain the lowest energy states presented in Fig. \ref{fig:gbenergies}.

Our simulations produce well-stitched GBs consisting of various types of defect rings.
At very small $\theta$ the GB dislocation cores are mostly
composed of well-separated $5|7$, $4|6$, or $4|10$ pairs,
with some examples ($5|7$ and $4|6$) shown in Fig. \ref{fig:gbenergies}b.
GB structures of connected $5|7$ arrays are found at larger misorientations
(Fig. \ref{fig:gbenergies}c), containing either a single type of homoelemental
neighboring, or the alternation of both types of opposite polarity
(i.e., alternating A-A and B-B) as observed in experiments of h-BN
\cite{GibbJACS13,LiNanoLett15}. Some new lowest-energy defect
structures are also obtained. These include $4|10$ pairs
comprising heteroelemental A-B neighboring (at $\theta=10.08^{\circ}$ of sym-zz),
$6|8$ or $4|6|8$ arrays (sym-zz $\theta=38.82^{\circ}$), and an interesting
case of GB between armchair and zigzag edges (asym $\theta=30^{\circ}$) showing as
a connected $5|7$ array mixed by $6|8$ ``fly-heads''. This ``fly-head''
configuration is similar to that obtained in graphene
\cite{LiuNanoLett10} as a result of the release of armchair-zigzag mismatch
stress, although here a new $7|6|8$ fly-head structure appears instead of
$7|5|7$ in graphene, given the energetically favorable heteroelemental-only
neighboring in the $6|8$ rings compared to the unfavorable homoelemental
one in $5|7$.

We have computed GB energies per unit length, $\gamma$, for all four types of GBs
setup. For each setup the lowest-energy results at $\theta$ from 0 to $60^{\circ}$
are given in Fig. \ref{fig:gbenergies}a (noting the full range of
$[0,120^{\circ}]$ is symmetric around $\theta=60^{\circ}$). All the $\gamma$
values calculated are within the range of previous first-principles DFT results
\cite{LiuACSNano12}. At small $\theta$ the $\gamma$ data are
well fitted to the Read-Shockley equation \cite{ReadShockley1950}
\begin{equation}
\gamma =\frac{b Y_2}{8\pi} \theta \left [ 1 + \ln(b/r_0) - \ln(2 \pi \theta ) \right ],
\label{eq:rs}
\end{equation}
where $Y_2$ is the 2D Young's modulus (set as 271 N/m for h-BN \cite{KudinPRB01}),
$b$ is the magnitude of the Burgers vector (assumed as
$b=a_0=2.51$ {\AA} for h-BN, for the shortest Burgers vector),
and $r_0$ is the dislocation core radius. The small-angle $\gamma$ values for
sym-zz and asym-zz cases are very close due to similar GB structures,
giving $r_0=0.54 \pm 0.01$ {\AA} from the fitting. For the sym-ac GBs, the
fitting yields $r_0=0.90 \pm 0.01$ {\AA} (in comparison, previous first-principles
calculations \cite{Yazyev10} gave $r_0=1.2$ {\AA} for graphene armchair GBs,
while no prior results for h-BN are available). In contrast, the asymmetric
tilting from armchair direction (asym-ac) leads to more sinuous configurations
with non-regular spacing of dislocation cores when compared to the sym-ac GBs
(see Fig. \ref{fig:gbenergies}b). This explains the higher
values and larger variation of small-angle $\gamma$ for asym-ac GBs. A different
value of $r_0=0.60 \pm 0.06$ {\AA} is also obtained from fitting.

\begin{figure}
\centerline{\includegraphics[width=0.5\textwidth]{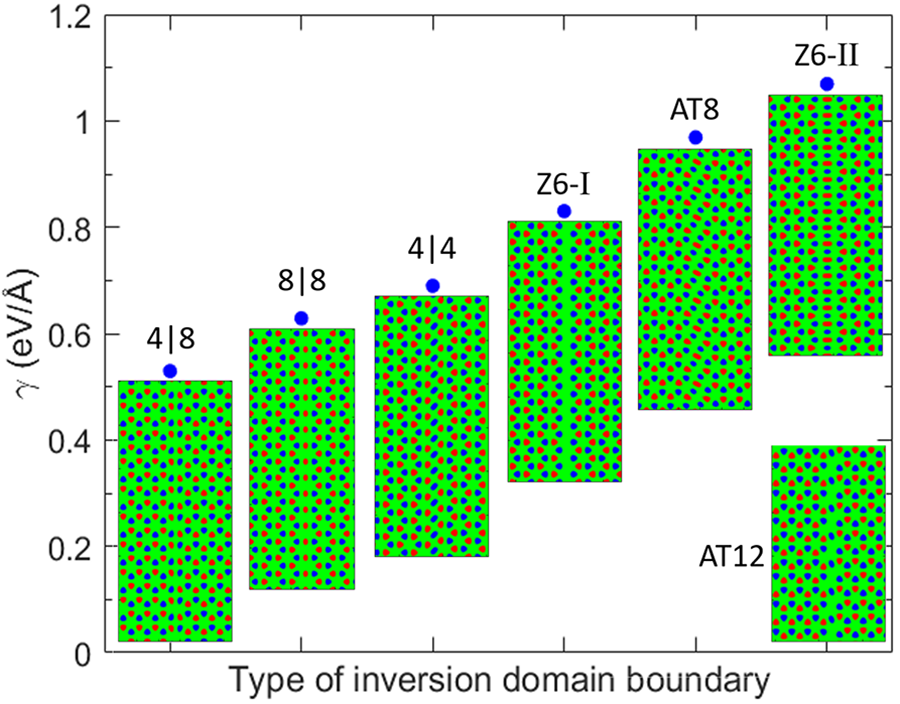}}
\caption{Energies and structures of various types of IDBs.}
\label{fig:idt}
\end{figure}

For $\theta$ approaching $60^{\circ}$, $4|8$ dislocation cores appear
more frequently in low-energy structures.
This square-octagon configuration, consisting only of
heteroelemental neighboring, is found to be the lowest energy state
of $60^{\circ}$ IDB, agreeing with the
first-principles result for h-BN \cite{LiuACSNano12}.
We also identify 6 other IDB structures that possess higher energy
(Fig. \ref{fig:idt}), including arrays of $8|8$, $4|4$, and tilted 6-membered
rings (Z6-I, equivalent to the absence of one A-atom column in $4|4$
structure) for zigzag GBs, arrays of tilted 8-membered (AT8) or tilted
12-membered (AT12) rings for armchair GBs, and a high-energy state composed
of compressed 6-6 pairs (Z6-II). The stable AT12 structure is obtained at
lower densities $n_{A0}=n_{B0}=-0.3$, but it is unstable (transforming
to $4|8$) at $n_{A0}=n_{B0}=-0.28$ used in all other GBs calculations.
Among them the $4|4$, $8|8$, and Z6-I structures, although not being found
in previous h-BN studies, have been obtained in some theoretical \cite{ZouNanoLett13}
and experimental \cite{EnyashinJPCC13,ZhouNanoLett13,LiuPRL14} work of
MoS$_2$ and MoSe$_2$ monolayers.

To identify the reliability and parameter dependence of our results, we have
conducted a large number of additional simulations particularly for IDBs,
with varying model parameters. All of them gave similar results for the
defect core structures and GB energy sequence. Details are described in
Supplemental Material.

To further examine the emergence and dynamics of $60^{\circ}$ boundaries,
we conduct a series of simulations of grain nucleation and growth from a
supersaturated homogeneous phase. The dynamics is modeled via
$\partial n_A / \partial t = - {\delta \mathcal{F}}/{\delta n_A} + \mu_A$ and
$\partial n_B / \partial t = - m_B ( {\delta \mathcal{F}}/{\delta n_B} - \mu_B )$,
where $\mu_{A(B)}$ is the chemical potential of A(B) component. We set $\mu_{A(B)}$
slightly above the equilibrium value to enable and control the grain growth via a
constant flux. Also $\mu_A=\mu_B$ is assumed, so that a single grain is of hexagonal
shape with both A- and B-terminated zigzag edges, consistent with some previous
experimental \cite{TayNanoLett14} and theoretical \cite{ZhaoPCCP15,ZhangNanoLett16}
results for h-BN.

\begin{figure}
\centerline{\includegraphics[width=0.5\textwidth]{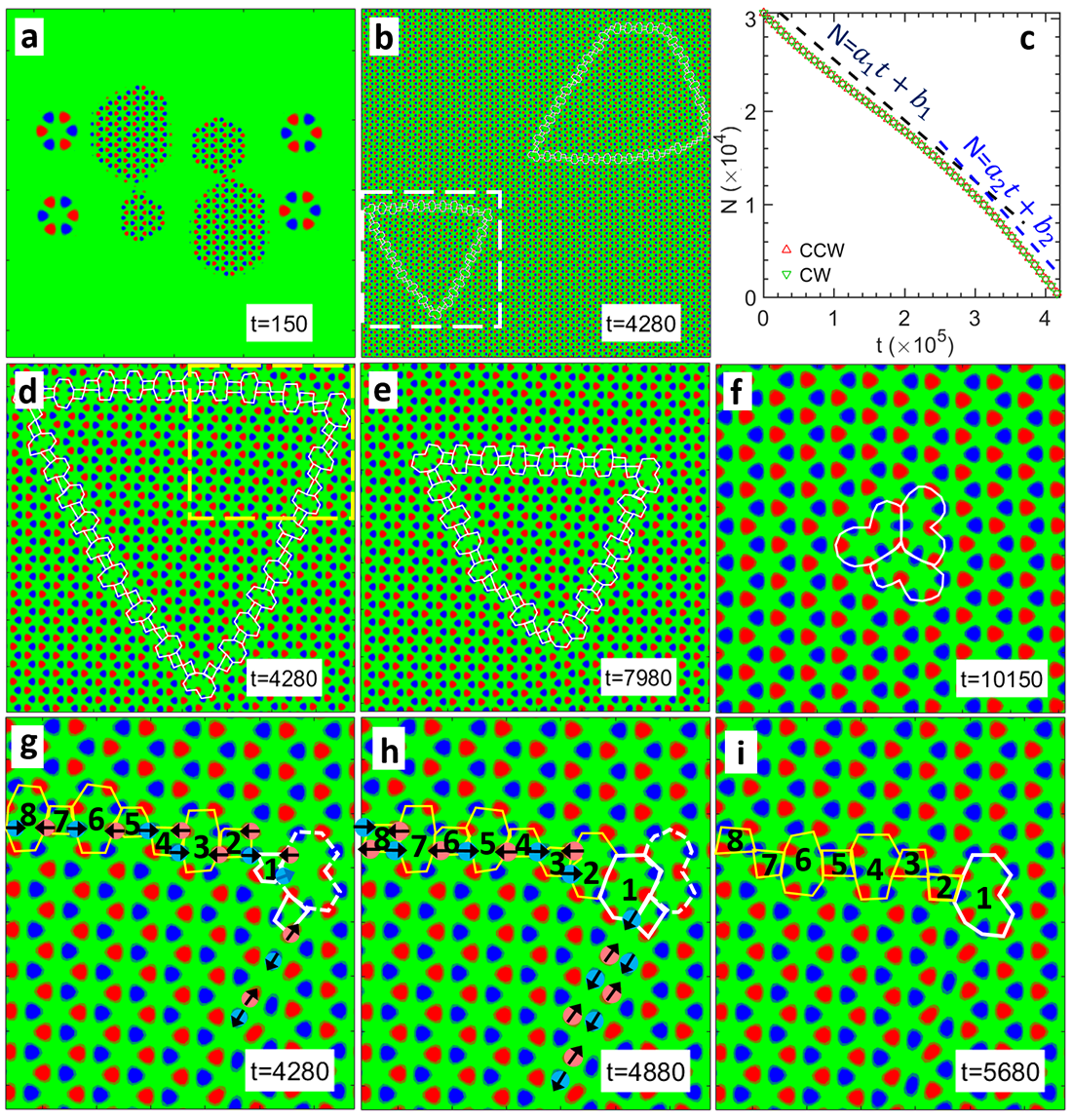}}
\caption{(a)-(f) Grain coalescence and inversion domain dynamics.
  (c) Atomic site number $N$ vs $t$, showing two regimes
  $N = -0.065 t + 3.045 \times 10^4$ and
  $N = -0.085 t + 3.612 \times 10^4$ via fitting. (d)-(e) Domain
  shrinking in the white boxed region of (b). (f) Transient of 3 merging
  heart-shaped defects before their annihilation. (g)-(i) Time evolution of
  collective atomic displacements in the yellow boxed corner of (d).}
\label{fig:dmofe}
\end{figure}

Figure \ref{fig:dmofe}a gives a typical simulation setup,
starting from four nuclei that are of two different sizes and $60^{\circ}$
misorientation. 
The nuclei evolve to grains of hexagon shape and grow individually
until they merge and form $60^{\circ}$ IDBs. This grain coalescence leads to
the embedding of polygon-shaped inversion domains within the large grain or
matrix and the subsequent domain shrinking (Fig. \ref{fig:dmofe}b).
The resulting IDBs can be composed not only of $4|8$ cores (the lowest-energy type),
but also of all other types of defect structure given in Fig. \ref{fig:idt}.
Many of them are transients and transform to lower-energy configurations
(mostly $4|8$, with some $8|8$ or $4|4$) during the evolution,
as seen in the supplemental movies \cite{SM}.

Figures \ref{fig:dmofe}d--f show the shrinking process of a typical triangle-shaped
inversion domain. The domain boundary lines consist of connected $4|8$ pairs, joined
at three junctions via decagon heart-shaped defect, with all defect cores containing
only heteroelemental neighboring. The 3 heart-shaped junctions will finally merge and
annihilate. The atomic evolution around a domain corner is detailed in
Fig. \ref{fig:dmofe}g--i, over a cycle associated with one atomic-step shrinkage.
Each cycle starts with the simultaneous displacement of atoms along two opposite
directions (indicated by the arrows in Fig. \ref{fig:dmofe}g)
inside each defect ring labeled from 1 to 8. This leads to the shape transformation
of defect rings at a later time (Fig. \ref{fig:dmofe}h), e.g., $4 \rightarrow 8$
transformation for ring 1, 2, 5, 7 and $8 \rightarrow 4$ for ring 3, 6, 8,
mediated by the rearrangement of corner decagon (white-dashed heart defect) to two
6-membered rings. Further atomic displacements inside rings 1-8 lead to further
ring shape transformation (Fig. \ref{fig:dmofe}i), e.g., $8 \rightarrow 4$ for
ring 2, 5, 7 and $4 \rightarrow 8$ for ring 4, 6, which are shifted downward
in comparison to their starting locations in Fig. \ref{fig:dmofe}g. Also the
heart-shaped junction is reconstructed at the location of ring 1 which, compared
to the junction at the beginning of this cycle (\ref{fig:dmofe}g, dashed heart),
is moved one-step inward of the triangle (of both 1 atomic-row downward and 1
atomic-column leftward). The whole triangle domain then shrinks one-step rigidly. 

Although similar heart-shaped defect (for h-BN \cite{CretuNanoLett14})
or triangle inversion domains (for MoSe$_2$ \cite{LinACSNano15,LehtinenACSNano15})
have been observed in experiments, they were generated by electron irradiation and
thus of qualitatively different structures and dynamics. The $4|8$ and
heart defects inside a single domain of irradiated h-BN flakes contained energetically
unfavorable homoelemental bonds \cite{CretuNanoLett14}, instead of heteroelemental-only
ones obtained here. In exfoliated or as-grown MoSe$_2$ sheets,
irradiation-induced Se vacancies gave rise to small triangle domains
with $4|4$ defects at both boundaries and junctions, and the expansion, but not
shrinking, of the domain is driven by the new vacancy creation and the triggered 
atomic motion \cite{LinACSNano15,LehtinenACSNano15}.

In contrast, here polygon-shape domains are formed through coalescing of $60^{\circ}$
misoriented grains during growth. The subsequent grain dynamics is characterized
by the collective atomic displacements and simultaneous shape transformation of
defect core rings along the GBs and junctions, and involves the glide of boundary
dislocations but not climb. Also no net translation, rotation or shear-induced
deformation has been observed, a scenario different from the Cahn-Taylor
shear driven mechanism with the coupling between normal and tangential motions
\cite{CahnActaMater04,*CahnActaMater06,McReynoldsActaMater16}. Such a mechanism
originates from the breaking and maintaining of lattice plane continuity across
a GB, and would not apply to the case of inversion domains for which the lattice
planes always remain continuous and a $60^{\circ}$ GB forms purely due to inversion
symmetry breaking in the binary lattice but not lattice sites mismatch.

During grain growth the characteristic grain size was historically expected
to follow a power law $L \propto t^\alpha$ (corresponding to $t^{2\alpha}$
for 2D grain area or atomic site number $N$), where the exponent $\alpha=1/2$
for the classical curvature-driven growth \cite{re:mullins56,*re:mullins89}
but is found to be $<1/2$ and dependent on system temperature, noise strength,
and bulk dissipation \cite{AdlandPRL13,BackofenActaMater14}
and also GBs roughness that affects grain growth stagnation \cite{HolmScience10}.
To quantify the collective dynamics identified here, we
simulate large systems of size 56.2 nm $\times$ 97.4 nm ($2 \times 10^5$ atomic
sites), with a single triangle inversion domain 
being initialized in the center and consisting of $4|8$ IDBs and heart-defect
junctions. As shown in Fig. \ref{fig:dmofe}c, two linear regimes of grain area
shrinkage can be identified, for which the crossover to the faster shrinking
rate occurs at a late time stage due to the defect interaction between GBs.
Although the exponent $\alpha=1/2$ is obtained here, the corresponding
dynamics of inversion domain is expected to be different from the conventional
curvature-driven mechanism, given the straight or weakly curved domain boundaries
and also the rigid and diffusionless boundary motion described above.
The domain shrinking can be significantly slowed down for some other types
of defects (e.g., a mixture of $4|8$, $4|4$, and $8|8$ cores),
leading to the existence of metastable polygon domains in the sample.
However, this still yields two linear regimes of decreasing $N$ (i.e.,
$\alpha=1/2$) before the stagnation of inversion domain, as shown in Fig. S6 of
Supplemental Material. More work is needed for further understanding this $\alpha=1/2$
behavior, which is important in identifying the growth mechanisms of binary 2D
materials that are different from those of traditional single-component or alloying
systems.

In summary, we have systematically studied the structural
and dynamical properties of 2D material systems at large spatial and temporal
scales that are not accessible to traditional atomistic methods.
Our results provide predictions and understanding of plasticity in binary 2D
materials, including some new defect structures of symmetrically and asymmetrically
tilted GBs and GB energies across the full range of misorientations.
We also predict a dynamic behavior of grain motion for inversion domains,
where the domain dynamics is governed by the collective
atomic displacements of the connected dislocation core pairs along the boundary
lines mediated via the heart-shaped defects at the junctions.
These findings as well as the modeling approach presented here provide further
insights for understanding the complex mechanisms of grain growth in binary 2D
materials.

\begin{acknowledgments}
  K.R.E. acknowledges support from the National Science Foundation
  under Grant No. DMR-1506634.
\end{acknowledgments}

\bibliography{hBN_GBs_References}

\begin{thebibliography}{44}%
\makeatletter
\providecommand \@ifxundefined [1]{%
 \@ifx{#1\undefined}
}%
\providecommand \@ifnum [1]{%
 \ifnum #1\expandafter \@firstoftwo
 \else \expandafter \@secondoftwo
 \fi
}%
\providecommand \@ifx [1]{%
 \ifx #1\expandafter \@firstoftwo
 \else \expandafter \@secondoftwo
 \fi
}%
\providecommand \natexlab [1]{#1}%
\providecommand \enquote  [1]{``#1''}%
\providecommand \bibnamefont  [1]{#1}%
\providecommand \bibfnamefont [1]{#1}%
\providecommand \citenamefont [1]{#1}%
\providecommand \href@noop [0]{\@secondoftwo}%
\providecommand \href [0]{\begingroup \@sanitize@url \@href}%
\providecommand \@href[1]{\@@startlink{#1}\@@href}%
\providecommand \@@href[1]{\endgroup#1\@@endlink}%
\providecommand \@sanitize@url [0]{\catcode `\\12\catcode `\$12\catcode
  `\&12\catcode `\#12\catcode `\^12\catcode `\_12\catcode `\%12\relax}%
\providecommand \@@startlink[1]{}%
\providecommand \@@endlink[0]{}%
\providecommand \url  [0]{\begingroup\@sanitize@url \@url }%
\providecommand \@url [1]{\endgroup\@href {#1}{\urlprefix }}%
\providecommand \urlprefix  [0]{URL }%
\providecommand \Eprint [0]{\href }%
\providecommand \doibase [0]{http://dx.doi.org/}%
\providecommand \selectlanguage [0]{\@gobble}%
\providecommand \bibinfo  [0]{\@secondoftwo}%
\providecommand \bibfield  [0]{\@secondoftwo}%
\providecommand \translation [1]{[#1]}%
\providecommand \BibitemOpen [0]{}%
\providecommand \bibitemStop [0]{}%
\providecommand \bibitemNoStop [0]{.\EOS\space}%
\providecommand \EOS [0]{\spacefactor3000\relax}%
\providecommand \BibitemShut  [1]{\csname bibitem#1\endcsname}%
\let\auto@bib@innerbib\@empty
\bibitem [{\citenamefont {Yazyev}\ and\ \citenamefont {Chen}(2014)}]{Yazyev14}%
  \BibitemOpen
  \bibfield  {author} {\bibinfo {author} {\bibfnamefont {O.~V.}\ \bibnamefont
  {Yazyev}}\ and\ \bibinfo {author} {\bibfnamefont {Y.~P.}\ \bibnamefont
  {Chen}},\ }\href@noop {} {\bibfield  {journal} {\bibinfo  {journal} {Nat.
  Nanotech.}\ }\textbf {\bibinfo {volume} {9}},\ \bibinfo {pages} {755}
  (\bibinfo {year} {2014})}\BibitemShut {NoStop}%
\bibitem [{\citenamefont {Yazyev}\ and\ \citenamefont
  {Louie}(2010)}]{Yazyev10}%
  \BibitemOpen
  \bibfield  {author} {\bibinfo {author} {\bibfnamefont {O.~V.}\ \bibnamefont
  {Yazyev}}\ and\ \bibinfo {author} {\bibfnamefont {S.~G.}\ \bibnamefont
  {Louie}},\ }\href@noop {} {\bibfield  {journal} {\bibinfo  {journal} {Phys.
  Rev. B}\ }\textbf {\bibinfo {volume} {81}},\ \bibinfo {pages} {195420}
  (\bibinfo {year} {2010})}\BibitemShut {NoStop}%
\bibitem [{\citenamefont {Liu}\ and\ \citenamefont
  {Yakobson}(2010)}]{LiuNanoLett10}%
  \BibitemOpen
  \bibfield  {author} {\bibinfo {author} {\bibfnamefont {Y.}~\bibnamefont
  {Liu}}\ and\ \bibinfo {author} {\bibfnamefont {B.~I.}\ \bibnamefont
  {Yakobson}},\ }\href@noop {} {\bibfield  {journal} {\bibinfo  {journal} {Nano
  Lett.}\ }\textbf {\bibinfo {volume} {10}},\ \bibinfo {pages} {2178} (\bibinfo
  {year} {2010})}\BibitemShut {NoStop}%
\bibitem [{\citenamefont {Liu}\ \emph {et~al.}(2012)\citenamefont {Liu},
  \citenamefont {Zou},\ and\ \citenamefont {Yakobson}}]{LiuACSNano12}%
  \BibitemOpen
  \bibfield  {author} {\bibinfo {author} {\bibfnamefont {Y.}~\bibnamefont
  {Liu}}, \bibinfo {author} {\bibfnamefont {X.}~\bibnamefont {Zou}}, \ and\
  \bibinfo {author} {\bibfnamefont {B.~I.}\ \bibnamefont {Yakobson}},\
  }\href@noop {} {\bibfield  {journal} {\bibinfo  {journal} {ACS Nano}\
  }\textbf {\bibinfo {volume} {6}},\ \bibinfo {pages} {7053} (\bibinfo {year}
  {2012})}\BibitemShut {NoStop}%
\bibitem [{\citenamefont {Zou}\ \emph {et~al.}(2013)\citenamefont {Zou},
  \citenamefont {Liu},\ and\ \citenamefont {Yakobson}}]{ZouNanoLett13}%
  \BibitemOpen
  \bibfield  {author} {\bibinfo {author} {\bibfnamefont {X.}~\bibnamefont
  {Zou}}, \bibinfo {author} {\bibfnamefont {Y.}~\bibnamefont {Liu}}, \ and\
  \bibinfo {author} {\bibfnamefont {B.~I.}\ \bibnamefont {Yakobson}},\
  }\href@noop {} {\bibfield  {journal} {\bibinfo  {journal} {Nano Lett.}\
  }\textbf {\bibinfo {volume} {13}},\ \bibinfo {pages} {253} (\bibinfo {year}
  {2013})}\BibitemShut {NoStop}%
\bibitem [{\citenamefont {Huang}\ \emph {et~al.}(2011)\citenamefont {Huang},
  \citenamefont {Ruiz-Vargas}, \citenamefont {{van der Zande}}, \citenamefont
  {Whitney}, \citenamefont {Levendorf}, \citenamefont {Kevek}, \citenamefont
  {Garg}, \citenamefont {Alden}, \citenamefont {Hustedt}, \citenamefont {Zhu},
  \citenamefont {Park}, \citenamefont {McEuen},\ and\ \citenamefont
  {Muller}}]{Huang11}%
  \BibitemOpen
  \bibfield  {author} {\bibinfo {author} {\bibfnamefont {P.~Y.}\ \bibnamefont
  {Huang}}, \bibinfo {author} {\bibfnamefont {C.~S.}\ \bibnamefont
  {Ruiz-Vargas}}, \bibinfo {author} {\bibfnamefont {A.~M.}\ \bibnamefont {{van
  der Zande}}}, \bibinfo {author} {\bibfnamefont {W.~S.}\ \bibnamefont
  {Whitney}}, \bibinfo {author} {\bibfnamefont {M.~P.}\ \bibnamefont
  {Levendorf}}, \bibinfo {author} {\bibfnamefont {J.~W.}\ \bibnamefont
  {Kevek}}, \bibinfo {author} {\bibfnamefont {S.}~\bibnamefont {Garg}},
  \bibinfo {author} {\bibfnamefont {J.~S.}\ \bibnamefont {Alden}}, \bibinfo
  {author} {\bibfnamefont {C.~J.}\ \bibnamefont {Hustedt}}, \bibinfo {author}
  {\bibfnamefont {Y.}~\bibnamefont {Zhu}}, \bibinfo {author} {\bibfnamefont
  {J.}~\bibnamefont {Park}}, \bibinfo {author} {\bibfnamefont {P.~L.}\
  \bibnamefont {McEuen}}, \ and\ \bibinfo {author} {\bibfnamefont {D.~A.}\
  \bibnamefont {Muller}},\ }\href@noop {} {\bibfield  {journal} {\bibinfo
  {journal} {Nature}\ }\textbf {\bibinfo {volume} {469}},\ \bibinfo {pages}
  {389} (\bibinfo {year} {2011})}\BibitemShut {NoStop}%
\bibitem [{\citenamefont {Gibb}\ \emph {et~al.}(2013)\citenamefont {Gibb},
  \citenamefont {Alem}, \citenamefont {Chen}, \citenamefont {Erickson},
  \citenamefont {Ciston}, \citenamefont {Gautam}, \citenamefont {Linck},\ and\
  \citenamefont {Zettl}}]{GibbJACS13}%
  \BibitemOpen
  \bibfield  {author} {\bibinfo {author} {\bibfnamefont {A.~L.}\ \bibnamefont
  {Gibb}}, \bibinfo {author} {\bibfnamefont {N.}~\bibnamefont {Alem}}, \bibinfo
  {author} {\bibfnamefont {J.-H.}\ \bibnamefont {Chen}}, \bibinfo {author}
  {\bibfnamefont {K.~J.}\ \bibnamefont {Erickson}}, \bibinfo {author}
  {\bibfnamefont {J.}~\bibnamefont {Ciston}}, \bibinfo {author} {\bibfnamefont
  {A.}~\bibnamefont {Gautam}}, \bibinfo {author} {\bibfnamefont
  {M.}~\bibnamefont {Linck}}, \ and\ \bibinfo {author} {\bibfnamefont
  {A.}~\bibnamefont {Zettl}},\ }\href@noop {} {\bibfield  {journal} {\bibinfo
  {journal} {J. Am. Chem. Soc.}\ }\textbf {\bibinfo {volume} {135}},\ \bibinfo
  {pages} {6758} (\bibinfo {year} {2013})}\BibitemShut {NoStop}%
\bibitem [{\citenamefont {Cretu}\ \emph {et~al.}(2014)\citenamefont {Cretu},
  \citenamefont {Lin},\ and\ \citenamefont {Suenaga}}]{CretuNanoLett14}%
  \BibitemOpen
  \bibfield  {author} {\bibinfo {author} {\bibfnamefont {O.}~\bibnamefont
  {Cretu}}, \bibinfo {author} {\bibfnamefont {Y.-C.}\ \bibnamefont {Lin}}, \
  and\ \bibinfo {author} {\bibfnamefont {K.}~\bibnamefont {Suenaga}},\
  }\href@noop {} {\bibfield  {journal} {\bibinfo  {journal} {Nano Lett.}\
  }\textbf {\bibinfo {volume} {14}},\ \bibinfo {pages} {1064} (\bibinfo {year}
  {2014})}\BibitemShut {NoStop}%
\bibitem [{\citenamefont {Li}\ \emph {et~al.}(2015)\citenamefont {Li},
  \citenamefont {Zou}, \citenamefont {Liu}, \citenamefont {Sun}, \citenamefont
  {Gao}, \citenamefont {Qi}, \citenamefont {Zhou}, \citenamefont {Yakobson},
  \citenamefont {Zhang},\ and\ \citenamefont {Liu}}]{LiNanoLett15}%
  \BibitemOpen
  \bibfield  {author} {\bibinfo {author} {\bibfnamefont {Q.}~\bibnamefont
  {Li}}, \bibinfo {author} {\bibfnamefont {X.}~\bibnamefont {Zou}}, \bibinfo
  {author} {\bibfnamefont {M.}~\bibnamefont {Liu}}, \bibinfo {author}
  {\bibfnamefont {J.}~\bibnamefont {Sun}}, \bibinfo {author} {\bibfnamefont
  {Y.}~\bibnamefont {Gao}}, \bibinfo {author} {\bibfnamefont {Y.}~\bibnamefont
  {Qi}}, \bibinfo {author} {\bibfnamefont {X.}~\bibnamefont {Zhou}}, \bibinfo
  {author} {\bibfnamefont {B.~I.}\ \bibnamefont {Yakobson}}, \bibinfo {author}
  {\bibfnamefont {Y.}~\bibnamefont {Zhang}}, \ and\ \bibinfo {author}
  {\bibfnamefont {Z.}~\bibnamefont {Liu}},\ }\href@noop {} {\bibfield
  {journal} {\bibinfo  {journal} {Nano Lett.}\ }\textbf {\bibinfo {volume}
  {15}},\ \bibinfo {pages} {5804} (\bibinfo {year} {2015})}\BibitemShut
  {NoStop}%
\bibitem [{\citenamefont {van~der Zande}\ \emph {et~al.}(2013)\citenamefont
  {van~der Zande}, \citenamefont {Huang}, \citenamefont {Chenet}, \citenamefont
  {Berkelbach}, \citenamefont {You}, \citenamefont {Lee}, \citenamefont
  {Heinz}, \citenamefont {Reichman}, \citenamefont {Muller},\ and\
  \citenamefont {Hone}}]{vanderZande13}%
  \BibitemOpen
  \bibfield  {author} {\bibinfo {author} {\bibfnamefont {A.~M.}\ \bibnamefont
  {van~der Zande}}, \bibinfo {author} {\bibfnamefont {P.~Y.}\ \bibnamefont
  {Huang}}, \bibinfo {author} {\bibfnamefont {D.~A.}\ \bibnamefont {Chenet}},
  \bibinfo {author} {\bibfnamefont {T.~C.}\ \bibnamefont {Berkelbach}},
  \bibinfo {author} {\bibfnamefont {Y.}~\bibnamefont {You}}, \bibinfo {author}
  {\bibfnamefont {G.-H.}\ \bibnamefont {Lee}}, \bibinfo {author} {\bibfnamefont
  {T.~F.}\ \bibnamefont {Heinz}}, \bibinfo {author} {\bibfnamefont {D.~R.}\
  \bibnamefont {Reichman}}, \bibinfo {author} {\bibfnamefont {D.~A.}\
  \bibnamefont {Muller}}, \ and\ \bibinfo {author} {\bibfnamefont {J.~C.}\
  \bibnamefont {Hone}},\ }\href@noop {} {\bibfield  {journal} {\bibinfo
  {journal} {Nat. Mater.}\ }\textbf {\bibinfo {volume} {12}},\ \bibinfo {pages}
  {554} (\bibinfo {year} {2013})}\BibitemShut {NoStop}%
\bibitem [{\citenamefont {Bj\"{o}rkman}\ \emph {et~al.}(2013)\citenamefont
  {Bj\"{o}rkman}, \citenamefont {Kurasch}, \citenamefont {Lehtinen},
  \citenamefont {Kotakoski}, \citenamefont {Yazyev}, \citenamefont
  {Srivastava}, \citenamefont {Skakalova}, \citenamefont {Smet}, \citenamefont
  {Kaiser},\ and\ \citenamefont {Krasheninnikov}}]{BjorkmanSciRep13}%
  \BibitemOpen
  \bibfield  {author} {\bibinfo {author} {\bibfnamefont {T.}~\bibnamefont
  {Bj\"{o}rkman}}, \bibinfo {author} {\bibfnamefont {S.}~\bibnamefont
  {Kurasch}}, \bibinfo {author} {\bibfnamefont {O.}~\bibnamefont {Lehtinen}},
  \bibinfo {author} {\bibfnamefont {J.}~\bibnamefont {Kotakoski}}, \bibinfo
  {author} {\bibfnamefont {O.~V.}\ \bibnamefont {Yazyev}}, \bibinfo {author}
  {\bibfnamefont {A.}~\bibnamefont {Srivastava}}, \bibinfo {author}
  {\bibfnamefont {V.}~\bibnamefont {Skakalova}}, \bibinfo {author}
  {\bibfnamefont {J.~H.}\ \bibnamefont {Smet}}, \bibinfo {author}
  {\bibfnamefont {U.}~\bibnamefont {Kaiser}}, \ and\ \bibinfo {author}
  {\bibfnamefont {A.~V.}\ \bibnamefont {Krasheninnikov}},\ }\href@noop {}
  {\bibfield  {journal} {\bibinfo  {journal} {Sci. Rep.}\ }\textbf {\bibinfo
  {volume} {3}},\ \bibinfo {pages} {3482} (\bibinfo {year} {2013})}\BibitemShut
  {NoStop}%
\bibitem [{\citenamefont {Zhou}\ \emph {et~al.}(2013)\citenamefont {Zhou},
  \citenamefont {Zou}, \citenamefont {Najmaei}, \citenamefont {Liu},
  \citenamefont {Shi}, \citenamefont {Kong}, \citenamefont {Lou}, \citenamefont
  {Ajayan}, \citenamefont {Yakobson},\ and\ \citenamefont
  {Idrobo}}]{ZhouNanoLett13}%
  \BibitemOpen
  \bibfield  {author} {\bibinfo {author} {\bibfnamefont {W.}~\bibnamefont
  {Zhou}}, \bibinfo {author} {\bibfnamefont {X.}~\bibnamefont {Zou}}, \bibinfo
  {author} {\bibfnamefont {S.}~\bibnamefont {Najmaei}}, \bibinfo {author}
  {\bibfnamefont {Z.}~\bibnamefont {Liu}}, \bibinfo {author} {\bibfnamefont
  {Y.}~\bibnamefont {Shi}}, \bibinfo {author} {\bibfnamefont {J.}~\bibnamefont
  {Kong}}, \bibinfo {author} {\bibfnamefont {J.}~\bibnamefont {Lou}}, \bibinfo
  {author} {\bibfnamefont {P.~M.}\ \bibnamefont {Ajayan}}, \bibinfo {author}
  {\bibfnamefont {B.~I.}\ \bibnamefont {Yakobson}}, \ and\ \bibinfo {author}
  {\bibfnamefont {J.-C.}\ \bibnamefont {Idrobo}},\ }\href@noop {} {\bibfield
  {journal} {\bibinfo  {journal} {Nano Lett.}\ }\textbf {\bibinfo {volume}
  {13}},\ \bibinfo {pages} {2615} (\bibinfo {year} {2013})}\BibitemShut
  {NoStop}%
\bibitem [{\citenamefont {Liu}\ \emph {et~al.}(2014)\citenamefont {Liu},
  \citenamefont {Jiao}, \citenamefont {Yang}, \citenamefont {Cai},
  \citenamefont {Wu}, \citenamefont {Ho}, \citenamefont {Gao}, \citenamefont
  {Jia}, \citenamefont {Wang}, \citenamefont {Fan}, \citenamefont {Yao},\ and\
  \citenamefont {Xie}}]{LiuPRL14}%
  \BibitemOpen
  \bibfield  {author} {\bibinfo {author} {\bibfnamefont {H.}~\bibnamefont
  {Liu}}, \bibinfo {author} {\bibfnamefont {L.}~\bibnamefont {Jiao}}, \bibinfo
  {author} {\bibfnamefont {F.}~\bibnamefont {Yang}}, \bibinfo {author}
  {\bibfnamefont {Y.}~\bibnamefont {Cai}}, \bibinfo {author} {\bibfnamefont
  {X.}~\bibnamefont {Wu}}, \bibinfo {author} {\bibfnamefont {W.}~\bibnamefont
  {Ho}}, \bibinfo {author} {\bibfnamefont {C.}~\bibnamefont {Gao}}, \bibinfo
  {author} {\bibfnamefont {J.}~\bibnamefont {Jia}}, \bibinfo {author}
  {\bibfnamefont {N.}~\bibnamefont {Wang}}, \bibinfo {author} {\bibfnamefont
  {H.}~\bibnamefont {Fan}}, \bibinfo {author} {\bibfnamefont {W.}~\bibnamefont
  {Yao}}, \ and\ \bibinfo {author} {\bibfnamefont {M.}~\bibnamefont {Xie}},\
  }\href@noop {} {\bibfield  {journal} {\bibinfo  {journal} {Phys. Rev. Lett.}\
  }\textbf {\bibinfo {volume} {113}},\ \bibinfo {pages} {066105} (\bibinfo
  {year} {2014})}\BibitemShut {NoStop}%
\bibitem [{\citenamefont {Lin}\ \emph {et~al.}(2015)\citenamefont {Lin},
  \citenamefont {Pantelides},\ and\ \citenamefont {Zhou}}]{LinACSNano15}%
  \BibitemOpen
  \bibfield  {author} {\bibinfo {author} {\bibfnamefont {J.}~\bibnamefont
  {Lin}}, \bibinfo {author} {\bibfnamefont {S.~T.}\ \bibnamefont {Pantelides}},
  \ and\ \bibinfo {author} {\bibfnamefont {W.}~\bibnamefont {Zhou}},\
  }\href@noop {} {\bibfield  {journal} {\bibinfo  {journal} {ACS Nano}\
  }\textbf {\bibinfo {volume} {9}},\ \bibinfo {pages} {5189} (\bibinfo {year}
  {2015})}\BibitemShut {NoStop}%
\bibitem [{\citenamefont {Lehtinen}\ \emph {et~al.}(2015)\citenamefont
  {Lehtinen}, \citenamefont {Komsa}, \citenamefont {Pulkin}, \citenamefont
  {Whitwick}, \citenamefont {Chen}, \citenamefont {Lehnert}, \citenamefont
  {Mohn}, \citenamefont {Yazyev}, \citenamefont {Kis}, \citenamefont {Kaiser},\
  and\ \citenamefont {Krasheninnikov}}]{LehtinenACSNano15}%
  \BibitemOpen
  \bibfield  {author} {\bibinfo {author} {\bibfnamefont {O.}~\bibnamefont
  {Lehtinen}}, \bibinfo {author} {\bibfnamefont {H.-P.}\ \bibnamefont {Komsa}},
  \bibinfo {author} {\bibfnamefont {A.}~\bibnamefont {Pulkin}}, \bibinfo
  {author} {\bibfnamefont {M.~B.}\ \bibnamefont {Whitwick}}, \bibinfo {author}
  {\bibfnamefont {M.-W.}\ \bibnamefont {Chen}}, \bibinfo {author}
  {\bibfnamefont {T.}~\bibnamefont {Lehnert}}, \bibinfo {author} {\bibfnamefont
  {M.~J.}\ \bibnamefont {Mohn}}, \bibinfo {author} {\bibfnamefont {O.~V.}\
  \bibnamefont {Yazyev}}, \bibinfo {author} {\bibfnamefont {A.}~\bibnamefont
  {Kis}}, \bibinfo {author} {\bibfnamefont {U.}~\bibnamefont {Kaiser}}, \ and\
  \bibinfo {author} {\bibfnamefont {A.~V.}\ \bibnamefont {Krasheninnikov}},\
  }\href@noop {} {\bibfield  {journal} {\bibinfo  {journal} {ACS Nano}\
  }\textbf {\bibinfo {volume} {9}},\ \bibinfo {pages} {3274} (\bibinfo {year}
  {2015})}\BibitemShut {NoStop}%
\bibitem [{\citenamefont {Elder}\ \emph {et~al.}(2002)\citenamefont {Elder},
  \citenamefont {Katakowski}, \citenamefont {Haataja},\ and\ \citenamefont
  {Grant}}]{Elder02}%
  \BibitemOpen
  \bibfield  {author} {\bibinfo {author} {\bibfnamefont {K.~R.}\ \bibnamefont
  {Elder}}, \bibinfo {author} {\bibfnamefont {M.}~\bibnamefont {Katakowski}},
  \bibinfo {author} {\bibfnamefont {M.}~\bibnamefont {Haataja}}, \ and\
  \bibinfo {author} {\bibfnamefont {M.}~\bibnamefont {Grant}},\ }\href@noop {}
  {\bibfield  {journal} {\bibinfo  {journal} {Phys. Rev. Lett.}\ }\textbf
  {\bibinfo {volume} {88}},\ \bibinfo {pages} {245701} (\bibinfo {year}
  {2002})}\BibitemShut {NoStop}%
\bibitem [{\citenamefont {Elder}\ and\ \citenamefont {Grant}(2004)}]{Elder04}%
  \BibitemOpen
  \bibfield  {author} {\bibinfo {author} {\bibfnamefont {K.~R.}\ \bibnamefont
  {Elder}}\ and\ \bibinfo {author} {\bibfnamefont {M.}~\bibnamefont {Grant}},\
  }\href@noop {} {\bibfield  {journal} {\bibinfo  {journal} {Phys. Rev. E}\
  }\textbf {\bibinfo {volume} {70}},\ \bibinfo {pages} {051605} (\bibinfo
  {year} {2004})}\BibitemShut {NoStop}%
\bibitem [{\citenamefont {Elder}\ \emph {et~al.}(2007)\citenamefont {Elder},
  \citenamefont {Provatas}, \citenamefont {Berry}, \citenamefont {Stefanovic},\
  and\ \citenamefont {Grant}}]{Elder07}%
  \BibitemOpen
  \bibfield  {author} {\bibinfo {author} {\bibfnamefont {K.~R.}\ \bibnamefont
  {Elder}}, \bibinfo {author} {\bibfnamefont {N.}~\bibnamefont {Provatas}},
  \bibinfo {author} {\bibfnamefont {J.}~\bibnamefont {Berry}}, \bibinfo
  {author} {\bibfnamefont {P.}~\bibnamefont {Stefanovic}}, \ and\ \bibinfo
  {author} {\bibfnamefont {M.}~\bibnamefont {Grant}},\ }\href@noop {}
  {\bibfield  {journal} {\bibinfo  {journal} {Phys. Rev. B}\ }\textbf {\bibinfo
  {volume} {75}},\ \bibinfo {pages} {064107} (\bibinfo {year}
  {2007})}\BibitemShut {NoStop}%
\bibitem [{\citenamefont {Emmerich}\ \emph {et~al.}(2012)\citenamefont
  {Emmerich}, \citenamefont {L{\"{o}}wen}, \citenamefont {Wittkowski},
  \citenamefont {Gruhn}, \citenamefont {T{\'{o}}th}, \citenamefont {Tegze},\
  and\ \citenamefont {Gr{\'{a}}n{\'{a}}sy}}]{Emmerich12}%
  \BibitemOpen
  \bibfield  {author} {\bibinfo {author} {\bibfnamefont {H.}~\bibnamefont
  {Emmerich}}, \bibinfo {author} {\bibfnamefont {H.}~\bibnamefont
  {L{\"{o}}wen}}, \bibinfo {author} {\bibfnamefont {R.}~\bibnamefont
  {Wittkowski}}, \bibinfo {author} {\bibfnamefont {T.}~\bibnamefont {Gruhn}},
  \bibinfo {author} {\bibfnamefont {G.~I.}\ \bibnamefont {T{\'{o}}th}},
  \bibinfo {author} {\bibfnamefont {G.}~\bibnamefont {Tegze}}, \ and\ \bibinfo
  {author} {\bibfnamefont {L.}~\bibnamefont {Gr{\'{a}}n{\'{a}}sy}},\
  }\href@noop {} {\bibfield  {journal} {\bibinfo  {journal} {Adv. Phys.}\
  }\textbf {\bibinfo {volume} {61}},\ \bibinfo {pages} {665} (\bibinfo {year}
  {2012})}\BibitemShut {NoStop}%
\bibitem [{\citenamefont {Mkhonta}\ \emph {et~al.}(2013)\citenamefont
  {Mkhonta}, \citenamefont {Elder},\ and\ \citenamefont {Huang}}]{Mkhonta13}%
  \BibitemOpen
  \bibfield  {author} {\bibinfo {author} {\bibfnamefont {S.~K.}\ \bibnamefont
  {Mkhonta}}, \bibinfo {author} {\bibfnamefont {K.~R.}\ \bibnamefont {Elder}},
  \ and\ \bibinfo {author} {\bibfnamefont {Z.-F.}\ \bibnamefont {Huang}},\
  }\href@noop {} {\bibfield  {journal} {\bibinfo  {journal} {Phys. Rev. Lett.}\
  }\textbf {\bibinfo {volume} {111}},\ \bibinfo {pages} {035501} (\bibinfo
  {year} {2013})}\BibitemShut {NoStop}%
\bibitem [{\citenamefont {Huang}\ and\ \citenamefont {Elder}(2008)}]{Huang08}%
  \BibitemOpen
  \bibfield  {author} {\bibinfo {author} {\bibfnamefont {Z.-F.}\ \bibnamefont
  {Huang}}\ and\ \bibinfo {author} {\bibfnamefont {K.~R.}\ \bibnamefont
  {Elder}},\ }\href@noop {} {\bibfield  {journal} {\bibinfo  {journal} {Phys.
  Rev. Lett.}\ }\textbf {\bibinfo {volume} {101}},\ \bibinfo {pages} {158701}
  (\bibinfo {year} {2008})}\BibitemShut {NoStop}%
\bibitem [{\citenamefont {Mellenthin}\ \emph {et~al.}(2008)\citenamefont
  {Mellenthin}, \citenamefont {Karma},\ and\ \citenamefont
  {Plapp}}]{Mellenthin08}%
  \BibitemOpen
  \bibfield  {author} {\bibinfo {author} {\bibfnamefont {J.}~\bibnamefont
  {Mellenthin}}, \bibinfo {author} {\bibfnamefont {A.}~\bibnamefont {Karma}}, \
  and\ \bibinfo {author} {\bibfnamefont {M.}~\bibnamefont {Plapp}},\
  }\href@noop {} {\bibfield  {journal} {\bibinfo  {journal} {Phys. Rev. B}\
  }\textbf {\bibinfo {volume} {78}},\ \bibinfo {pages} {184110} (\bibinfo
  {year} {2008})}\BibitemShut {NoStop}%
\bibitem [{\citenamefont {T\'oth}\ \emph {et~al.}(2011)\citenamefont {T\'oth},
  \citenamefont {Pusztai}, \citenamefont {Tegze}, \citenamefont {T\'oth},\ and\
  \citenamefont {Gr\'an\'asy}}]{TothPRL11}%
  \BibitemOpen
  \bibfield  {author} {\bibinfo {author} {\bibfnamefont {G.~I.}\ \bibnamefont
  {T\'oth}}, \bibinfo {author} {\bibfnamefont {T.}~\bibnamefont {Pusztai}},
  \bibinfo {author} {\bibfnamefont {G.}~\bibnamefont {Tegze}}, \bibinfo
  {author} {\bibfnamefont {G.}~\bibnamefont {T\'oth}}, \ and\ \bibinfo {author}
  {\bibfnamefont {L.}~\bibnamefont {Gr\'an\'asy}},\ }\href@noop {} {\bibfield
  {journal} {\bibinfo  {journal} {Phys. Rev. Lett.}\ }\textbf {\bibinfo
  {volume} {107}},\ \bibinfo {pages} {175702} (\bibinfo {year}
  {2011})}\BibitemShut {NoStop}%
\bibitem [{\citenamefont {Schwalbach}\ \emph {et~al.}(2013)\citenamefont
  {Schwalbach}, \citenamefont {Warren}, \citenamefont {Wu},\ and\ \citenamefont
  {Voorhees}}]{SchwalbachPRE13}%
  \BibitemOpen
  \bibfield  {author} {\bibinfo {author} {\bibfnamefont {E.~J.}\ \bibnamefont
  {Schwalbach}}, \bibinfo {author} {\bibfnamefont {J.~A.}\ \bibnamefont
  {Warren}}, \bibinfo {author} {\bibfnamefont {K.-A.}\ \bibnamefont {Wu}}, \
  and\ \bibinfo {author} {\bibfnamefont {P.~W.}\ \bibnamefont {Voorhees}},\
  }\href@noop {} {\bibfield  {journal} {\bibinfo  {journal} {Phys. Rev. E}\
  }\textbf {\bibinfo {volume} {88}},\ \bibinfo {pages} {023306} (\bibinfo
  {year} {2013})}\BibitemShut {NoStop}%
\bibitem [{\citenamefont {Adland}\ \emph {et~al.}(2013)\citenamefont {Adland},
  \citenamefont {Xu},\ and\ \citenamefont {Karma}}]{AdlandPRL13}%
  \BibitemOpen
  \bibfield  {author} {\bibinfo {author} {\bibfnamefont {A.}~\bibnamefont
  {Adland}}, \bibinfo {author} {\bibfnamefont {Y.}~\bibnamefont {Xu}}, \ and\
  \bibinfo {author} {\bibfnamefont {A.}~\bibnamefont {Karma}},\ }\href@noop {}
  {\bibfield  {journal} {\bibinfo  {journal} {Phys. Rev. Lett.}\ }\textbf
  {\bibinfo {volume} {110}},\ \bibinfo {pages} {265504} (\bibinfo {year}
  {2013})}\BibitemShut {NoStop}%
\bibitem [{\citenamefont {Ofori-Opoku}\ \emph {et~al.}(2013)\citenamefont
  {Ofori-Opoku}, \citenamefont {Fallah}, \citenamefont {Greenwood},
  \citenamefont {Esmaeili},\ and\ \citenamefont {Provatas}}]{OpokuPRB13}%
  \BibitemOpen
  \bibfield  {author} {\bibinfo {author} {\bibfnamefont {N.}~\bibnamefont
  {Ofori-Opoku}}, \bibinfo {author} {\bibfnamefont {V.}~\bibnamefont {Fallah}},
  \bibinfo {author} {\bibfnamefont {M.}~\bibnamefont {Greenwood}}, \bibinfo
  {author} {\bibfnamefont {S.}~\bibnamefont {Esmaeili}}, \ and\ \bibinfo
  {author} {\bibfnamefont {N.}~\bibnamefont {Provatas}},\ }\href@noop {}
  {\bibfield  {journal} {\bibinfo  {journal} {Phys. Rev. B}\ }\textbf {\bibinfo
  {volume} {87}},\ \bibinfo {pages} {134105} (\bibinfo {year}
  {2013})}\BibitemShut {NoStop}%
\bibitem [{\citenamefont {Backofen}\ \emph {et~al.}(2014)\citenamefont
  {Backofen}, \citenamefont {Barmak}, \citenamefont {Elder},\ and\
  \citenamefont {Voigt}}]{BackofenActaMater14}%
  \BibitemOpen
  \bibfield  {author} {\bibinfo {author} {\bibfnamefont {R.}~\bibnamefont
  {Backofen}}, \bibinfo {author} {\bibfnamefont {K.}~\bibnamefont {Barmak}},
  \bibinfo {author} {\bibfnamefont {K.}~\bibnamefont {Elder}}, \ and\ \bibinfo
  {author} {\bibfnamefont {A.}~\bibnamefont {Voigt}},\ }\href@noop {}
  {\bibfield  {journal} {\bibinfo  {journal} {Acta Mater.}\ }\textbf {\bibinfo
  {volume} {64}},\ \bibinfo {pages} {72} (\bibinfo {year} {2014})}\BibitemShut
  {NoStop}%
\bibitem [{\citenamefont {Hirvonen}\ \emph {et~al.}(2016)\citenamefont
  {Hirvonen}, \citenamefont {Ervasti}, \citenamefont {Fan}, \citenamefont
  {Jalalvand}, \citenamefont {Seymour}, \citenamefont {Vaez~Allaei},
  \citenamefont {Provatas}, \citenamefont {Harju}, \citenamefont {Elder},\ and\
  \citenamefont {Ala-Nissila}}]{HirvonenPRB16}%
  \BibitemOpen
  \bibfield  {author} {\bibinfo {author} {\bibfnamefont {P.}~\bibnamefont
  {Hirvonen}}, \bibinfo {author} {\bibfnamefont {M.~M.}\ \bibnamefont
  {Ervasti}}, \bibinfo {author} {\bibfnamefont {Z.}~\bibnamefont {Fan}},
  \bibinfo {author} {\bibfnamefont {M.}~\bibnamefont {Jalalvand}}, \bibinfo
  {author} {\bibfnamefont {M.}~\bibnamefont {Seymour}}, \bibinfo {author}
  {\bibfnamefont {S.~M.}\ \bibnamefont {Vaez~Allaei}}, \bibinfo {author}
  {\bibfnamefont {N.}~\bibnamefont {Provatas}}, \bibinfo {author}
  {\bibfnamefont {A.}~\bibnamefont {Harju}}, \bibinfo {author} {\bibfnamefont
  {K.~R.}\ \bibnamefont {Elder}}, \ and\ \bibinfo {author} {\bibfnamefont
  {T.}~\bibnamefont {Ala-Nissila}},\ }\href@noop {} {\bibfield  {journal}
  {\bibinfo  {journal} {Phys. Rev. B}\ }\textbf {\bibinfo {volume} {94}},\
  \bibinfo {pages} {035414} (\bibinfo {year} {2016})}\BibitemShut {NoStop}%
\bibitem [{\citenamefont {Mkhonta}\ \emph {et~al.}(2016)\citenamefont
  {Mkhonta}, \citenamefont {Elder},\ and\ \citenamefont {Huang}}]{Mkhonta16}%
  \BibitemOpen
  \bibfield  {author} {\bibinfo {author} {\bibfnamefont {S.~K.}\ \bibnamefont
  {Mkhonta}}, \bibinfo {author} {\bibfnamefont {K.~R.}\ \bibnamefont {Elder}},
  \ and\ \bibinfo {author} {\bibfnamefont {Z.-F.}\ \bibnamefont {Huang}},\
  }\href@noop {} {\bibfield  {journal} {\bibinfo  {journal} {Phys. Rev. Lett.}\
  }\textbf {\bibinfo {volume} {116}},\ \bibinfo {pages} {205502} (\bibinfo
  {year} {2016})}\BibitemShut {NoStop}%
\bibitem [{\citenamefont {Huang}(2016)}]{Huang16}%
  \BibitemOpen
  \bibfield  {author} {\bibinfo {author} {\bibfnamefont {Z.-F.}\ \bibnamefont
  {Huang}},\ }\href@noop {} {\bibfield  {journal} {\bibinfo  {journal} {Phys.
  Rev. E}\ }\textbf {\bibinfo {volume} {93}},\ \bibinfo {pages} {022803}
  (\bibinfo {year} {2016})}\BibitemShut {NoStop}%
\bibitem [{SM()}]{SM}%
  \BibitemOpen
  \href@noop {} {}\bibinfo {note} {See Supplemental Material for model
  description, sample stability and phase diagrams, model parameterization,
  finite size effects in GB energy computation, additional GB structures and
  results of grain dynamics, and movies of grain coalescence and inversion
  domain evolution.}\BibitemShut {Stop}%
\bibitem [{\citenamefont {Huang}\ \emph {et~al.}(2010)\citenamefont {Huang},
  \citenamefont {Elder},\ and\ \citenamefont {Provatas}}]{Huang10}%
  \BibitemOpen
  \bibfield  {author} {\bibinfo {author} {\bibfnamefont {Z.-F.}\ \bibnamefont
  {Huang}}, \bibinfo {author} {\bibfnamefont {K.~R.}\ \bibnamefont {Elder}}, \
  and\ \bibinfo {author} {\bibfnamefont {N.}~\bibnamefont {Provatas}},\
  }\href@noop {} {\bibfield  {journal} {\bibinfo  {journal} {Phys. Rev. E}\
  }\textbf {\bibinfo {volume} {82}},\ \bibinfo {pages} {021605} (\bibinfo
  {year} {2010})}\BibitemShut {NoStop}%
\bibitem [{\citenamefont {Kudin}\ \emph {et~al.}(2001)\citenamefont {Kudin},
  \citenamefont {Scuseria},\ and\ \citenamefont {Yakobson}}]{KudinPRB01}%
  \BibitemOpen
  \bibfield  {author} {\bibinfo {author} {\bibfnamefont {K.~N.}\ \bibnamefont
  {Kudin}}, \bibinfo {author} {\bibfnamefont {G.~E.}\ \bibnamefont {Scuseria}},
  \ and\ \bibinfo {author} {\bibfnamefont {B.~I.}\ \bibnamefont {Yakobson}},\
  }\href@noop {} {\bibfield  {journal} {\bibinfo  {journal} {Phys. Rev. B}\
  }\textbf {\bibinfo {volume} {64}},\ \bibinfo {pages} {235406} (\bibinfo
  {year} {2001})}\BibitemShut {NoStop}%
\bibitem [{\citenamefont {Read}\ and\ \citenamefont
  {Shockley}(1950)}]{ReadShockley1950}%
  \BibitemOpen
  \bibfield  {author} {\bibinfo {author} {\bibfnamefont {W.~T.}\ \bibnamefont
  {Read}}\ and\ \bibinfo {author} {\bibfnamefont {W.}~\bibnamefont
  {Shockley}},\ }\href@noop {} {\bibfield  {journal} {\bibinfo  {journal}
  {Phys. Rev.}\ }\textbf {\bibinfo {volume} {78}},\ \bibinfo {pages} {275}
  (\bibinfo {year} {1950})}\BibitemShut {NoStop}%
\bibitem [{\citenamefont {Enyashin}\ \emph {et~al.}(2013)\citenamefont
  {Enyashin}, \citenamefont {Bar-Sadan}, \citenamefont {Houben},\ and\
  \citenamefont {Seifert}}]{EnyashinJPCC13}%
  \BibitemOpen
  \bibfield  {author} {\bibinfo {author} {\bibfnamefont {A.}~\bibnamefont
  {Enyashin}}, \bibinfo {author} {\bibfnamefont {M.}~\bibnamefont {Bar-Sadan}},
  \bibinfo {author} {\bibfnamefont {L.}~\bibnamefont {Houben}}, \ and\ \bibinfo
  {author} {\bibfnamefont {G.}~\bibnamefont {Seifert}},\ }\href@noop {}
  {\bibfield  {journal} {\bibinfo  {journal} {J. Phys. Chem. C}\ }\textbf
  {\bibinfo {volume} {117}},\ \bibinfo {pages} {10842} (\bibinfo {year}
  {2013})}\BibitemShut {NoStop}%
\bibitem [{\citenamefont {Tay}\ \emph {et~al.}(2014)\citenamefont {Tay},
  \citenamefont {Griep}, \citenamefont {Mallick}, \citenamefont {Tsang},
  \citenamefont {Singh}, \citenamefont {Tumlin}, \citenamefont {Teo},\ and\
  \citenamefont {Karna}}]{TayNanoLett14}%
  \BibitemOpen
  \bibfield  {author} {\bibinfo {author} {\bibfnamefont {R.~Y.}\ \bibnamefont
  {Tay}}, \bibinfo {author} {\bibfnamefont {M.~H.}\ \bibnamefont {Griep}},
  \bibinfo {author} {\bibfnamefont {G.}~\bibnamefont {Mallick}}, \bibinfo
  {author} {\bibfnamefont {S.~H.}\ \bibnamefont {Tsang}}, \bibinfo {author}
  {\bibfnamefont {R.~S.}\ \bibnamefont {Singh}}, \bibinfo {author}
  {\bibfnamefont {T.}~\bibnamefont {Tumlin}}, \bibinfo {author} {\bibfnamefont
  {E.~H.~T.}\ \bibnamefont {Teo}}, \ and\ \bibinfo {author} {\bibfnamefont
  {S.~P.}\ \bibnamefont {Karna}},\ }\href@noop {} {\bibfield  {journal}
  {\bibinfo  {journal} {Nano Lett.}\ }\textbf {\bibinfo {volume} {14}},\
  \bibinfo {pages} {839} (\bibinfo {year} {2014})}\BibitemShut {NoStop}%
\bibitem [{\citenamefont {Zhao}\ \emph {et~al.}(2015)\citenamefont {Zhao},
  \citenamefont {Li}, \citenamefont {Liu}, \citenamefont {Liu},\ and\
  \citenamefont {Ding}}]{ZhaoPCCP15}%
  \BibitemOpen
  \bibfield  {author} {\bibinfo {author} {\bibfnamefont {R.}~\bibnamefont
  {Zhao}}, \bibinfo {author} {\bibfnamefont {F.}~\bibnamefont {Li}}, \bibinfo
  {author} {\bibfnamefont {Z.}~\bibnamefont {Liu}}, \bibinfo {author}
  {\bibfnamefont {Z.}~\bibnamefont {Liu}}, \ and\ \bibinfo {author}
  {\bibfnamefont {F.}~\bibnamefont {Ding}},\ }\href@noop {} {\bibfield
  {journal} {\bibinfo  {journal} {Phys. Chem. Chem. Phys.}\ }\textbf {\bibinfo
  {volume} {17}},\ \bibinfo {pages} {29327} (\bibinfo {year}
  {2015})}\BibitemShut {NoStop}%
\bibitem [{\citenamefont {Zhang}\ \emph {et~al.}(2016)\citenamefont {Zhang},
  \citenamefont {Liu}, \citenamefont {Yang},\ and\ \citenamefont
  {Yakobson}}]{ZhangNanoLett16}%
  \BibitemOpen
  \bibfield  {author} {\bibinfo {author} {\bibfnamefont {Z.}~\bibnamefont
  {Zhang}}, \bibinfo {author} {\bibfnamefont {Y.}~\bibnamefont {Liu}}, \bibinfo
  {author} {\bibfnamefont {Y.}~\bibnamefont {Yang}}, \ and\ \bibinfo {author}
  {\bibfnamefont {B.~I.}\ \bibnamefont {Yakobson}},\ }\href@noop {} {\bibfield
  {journal} {\bibinfo  {journal} {Nano Lett.}\ }\textbf {\bibinfo {volume}
  {16}},\ \bibinfo {pages} {1398} (\bibinfo {year} {2016})}\BibitemShut
  {NoStop}%
\bibitem [{\citenamefont {Cahn}\ and\ \citenamefont
  {Taylor}(2004)}]{CahnActaMater04}%
  \BibitemOpen
  \bibfield  {author} {\bibinfo {author} {\bibfnamefont {J.~W.}\ \bibnamefont
  {Cahn}}\ and\ \bibinfo {author} {\bibfnamefont {J.~E.}\ \bibnamefont
  {Taylor}},\ }\href@noop {} {\bibfield  {journal} {\bibinfo  {journal} {Acta
  Mater.}\ }\textbf {\bibinfo {volume} {52}},\ \bibinfo {pages} {4887}
  (\bibinfo {year} {2004})}\BibitemShut {NoStop}%
\bibitem [{\citenamefont {Cahn}\ \emph {et~al.}(2006)\citenamefont {Cahn},
  \citenamefont {Mishin},\ and\ \citenamefont {Suzuki}}]{CahnActaMater06}%
  \BibitemOpen
  \bibfield  {author} {\bibinfo {author} {\bibfnamefont {J.~W.}\ \bibnamefont
  {Cahn}}, \bibinfo {author} {\bibfnamefont {Y.}~\bibnamefont {Mishin}}, \ and\
  \bibinfo {author} {\bibfnamefont {A.}~\bibnamefont {Suzuki}},\ }\href@noop {}
  {\bibfield  {journal} {\bibinfo  {journal} {Acta Mater.}\ }\textbf {\bibinfo
  {volume} {54}},\ \bibinfo {pages} {4953} (\bibinfo {year}
  {2006})}\BibitemShut {NoStop}%
\bibitem [{\citenamefont {McReynolds}\ \emph {et~al.}(2016)\citenamefont
  {McReynolds}, \citenamefont {Wu},\ and\ \citenamefont
  {Voorhees}}]{McReynoldsActaMater16}%
  \BibitemOpen
  \bibfield  {author} {\bibinfo {author} {\bibfnamefont {K.}~\bibnamefont
  {McReynolds}}, \bibinfo {author} {\bibfnamefont {K.-A.}\ \bibnamefont {Wu}},
  \ and\ \bibinfo {author} {\bibfnamefont {P.}~\bibnamefont {Voorhees}},\
  }\href@noop {} {\bibfield  {journal} {\bibinfo  {journal} {Acta Mater.}\
  }\textbf {\bibinfo {volume} {120}},\ \bibinfo {pages} {264} (\bibinfo {year}
  {2016})}\BibitemShut {NoStop}%
\bibitem [{\citenamefont {Mullins}(1956)}]{re:mullins56}%
  \BibitemOpen
  \bibfield  {author} {\bibinfo {author} {\bibfnamefont {W.~W.}\ \bibnamefont
  {Mullins}},\ }\href@noop {} {\bibfield  {journal} {\bibinfo  {journal} {J.
  Appl. Phys.}\ }\textbf {\bibinfo {volume} {27}},\ \bibinfo {pages} {900}
  (\bibinfo {year} {1956})}\BibitemShut {NoStop}%
\bibitem [{\citenamefont {Mullins}\ and\ \citenamefont
  {{Vi\~nals}}(1989)}]{re:mullins89}%
  \BibitemOpen
  \bibfield  {author} {\bibinfo {author} {\bibfnamefont {W.~W.}\ \bibnamefont
  {Mullins}}\ and\ \bibinfo {author} {\bibfnamefont {J.}~\bibnamefont
  {{Vi\~nals}}},\ }\href@noop {} {\bibfield  {journal} {\bibinfo  {journal}
  {Acta Metall.}\ }\textbf {\bibinfo {volume} {37}},\ \bibinfo {pages} {991}
  (\bibinfo {year} {1989})}\BibitemShut {NoStop}%
\bibitem [{\citenamefont {Holm}\ and\ \citenamefont
  {Foiles}(2010)}]{HolmScience10}%
  \BibitemOpen
  \bibfield  {author} {\bibinfo {author} {\bibfnamefont {E.~A.}\ \bibnamefont
  {Holm}}\ and\ \bibinfo {author} {\bibfnamefont {S.~M.}\ \bibnamefont
  {Foiles}},\ }\href@noop {} {\bibfield  {journal} {\bibinfo  {journal}
  {Science}\ }\textbf {\bibinfo {volume} {328}},\ \bibinfo {pages} {1138}
  (\bibinfo {year} {2010})}\BibitemShut {NoStop}%
\end{thebibliography}%

\end{document}